\documentclass[9pt, conference]{IEEEtran}
\IEEEoverridecommandlockouts
\usepackage{cite}
\usepackage{amsmath,amssymb,amsfonts}
\usepackage{algorithmic}
\usepackage{graphicx}
\usepackage{textcomp}
\usepackage{xcolor}
\def\BibTeX{{\rm B\kern-.05em{\sc i\kern-.025em b}\kern-.08em
    T\kern-.1667em\lower.7ex\hbox{E}\kern-.125emX}}

\usepackage{hyperref}
\usepackage{pifont}
\usepackage{multirow}
\usepackage{float}
\usepackage{cleveref}
\usepackage{xspace}

\usepackage{cjhebrew}
\usepackage{booktabs}
\usepackage{color}
\usepackage{etoolbox}
\usepackage{amsmath}

\newcommand{\newpara}[1]{\vspace{0.25cm} \noindent {\bf #1}}
\newcommand{\bench}{\textsc{SALMon}\xspace}

\title{Salmon: A Suite for Acoustic Language Model Evaluation}
\author{\IEEEauthorblockN{Gallil Maimon$^\dag$}
\IEEEauthorblockA{\textit{School of Computer Science \& Eng.} \\
\textit{Hebrew University of Jerusalem}\\
\textit{Jerusalem, Israel}}
\and
\IEEEauthorblockN{Amit Roth$^\dag$}
\IEEEauthorblockA{\textit{School of Computer Science \& Eng.} \\
\textit{Hebrew University of Jerusalem}\\
\textit{Jerusalem, Israel}}
\and
\IEEEauthorblockN{Yossi Adi\thanks{\textdagger Equal contribution.}}
\IEEEauthorblockA{\textit{School of Computer Science \& Eng.} \\
\textit{Hebrew University of Jerusalem}\\
\textit{Jerusalem, Israel}}\\
}

\begin{document}

\maketitle

\begin{abstract}
Speech language models have recently demonstrated great potential as universal speech processing systems. Such models have the ability to model the rich acoustic information existing in audio signals, beyond spoken content, such as emotion, background noise, etc. Despite this, evaluation benchmarks which evaluate awareness to a wide range of acoustic aspects, are lacking. To help bridge this gap, we introduce \bench, a novel evaluation suite encompassing background noise, sentiment, speaker identity and room impulse response. The proposed benchmarks both evaluate the consistency of the inspected element and how much it matches the spoken text. We follow a modelling based approach, measuring whether a model gives correct samples higher scores than incorrect ones. This approach makes the benchmark fast to compute even for large models. We evaluated several speech language models on \bench, thus highlighting the strengths and weaknesses of each evaluated method. We make the code and data publicly available at - \href{https://pages.cs.huji.ac.il/adiyoss-lab/salmon/}{pages.cs.huji.ac.il/adiyoss-lab/salmon/}.
\end{abstract}

\begin{IEEEkeywords}
Speech Language Models, Acoustic Modelling
\end{IEEEkeywords}

\section{Introduction} \label{sec:intro}
Speech Language Models (SLM) have recently gained great popularity as universal speech processing systems~\cite{chang2024speechprompt}. Since early approaches \cite{gslm}, many improvements have been achieved through scaling \cite{twist,scaling_laws}, using text LMs \cite{twist,spiritlm,maiti2024voxtlm,lu2024desta, salmonn}, dialogue modelling~\cite{nguyen2023generative}, and aligning with human preferences~\cite{zhang2024speechalign, fathullah2024audiochatllama}. Recently, such models have demonstrated impressive capabilities in real-time modelling of interactive conversations~\cite{zhang2024beyond, tanzer2024modeling, wang2024full, ma2024language}. These improvements manifested in higher semantic coherence of the generated speech, as shown by various text-based metrics that measure syntax and grammar~\cite{zero_resource}, or semantics~\cite{twist}, with little emphasis on other acoustic characteristics.

However, audio has many aspects other than spoken content such as sentiment, speaker identity, background noise, reverberation etc. For instance, the phrase ``Yeah, I'm sure'' could be sarcastic indicating lack of belief or sincere and sympathetic, depending on the intonation. Likewise, a request for a song recommendation to fit a certain atmosphere might elicit highly different responses if the background noise and accent indicate a beach in Mexico or a formal wedding in England. 

While some SLMs attempt to also model prosodic features \cite{pgslm, spiritlm}, evaluation of their ability to model different aspects is still lacking. The authors in~\cite{prosaudit} introduced ProsAudit which evaluates models' ability to assign higher likelihood to utterances with natural pauses. In addition, in \cite{spiritlm} the authors presented STSP which uses a pre-trained sentiment classifier to evaluate whether the generated speech sentiment is consistent with the prompt. Although they are highly valuable, these works evaluate a single aspect only.

\begin{figure}[t!]
\centering
\includegraphics[width=1.0\linewidth]{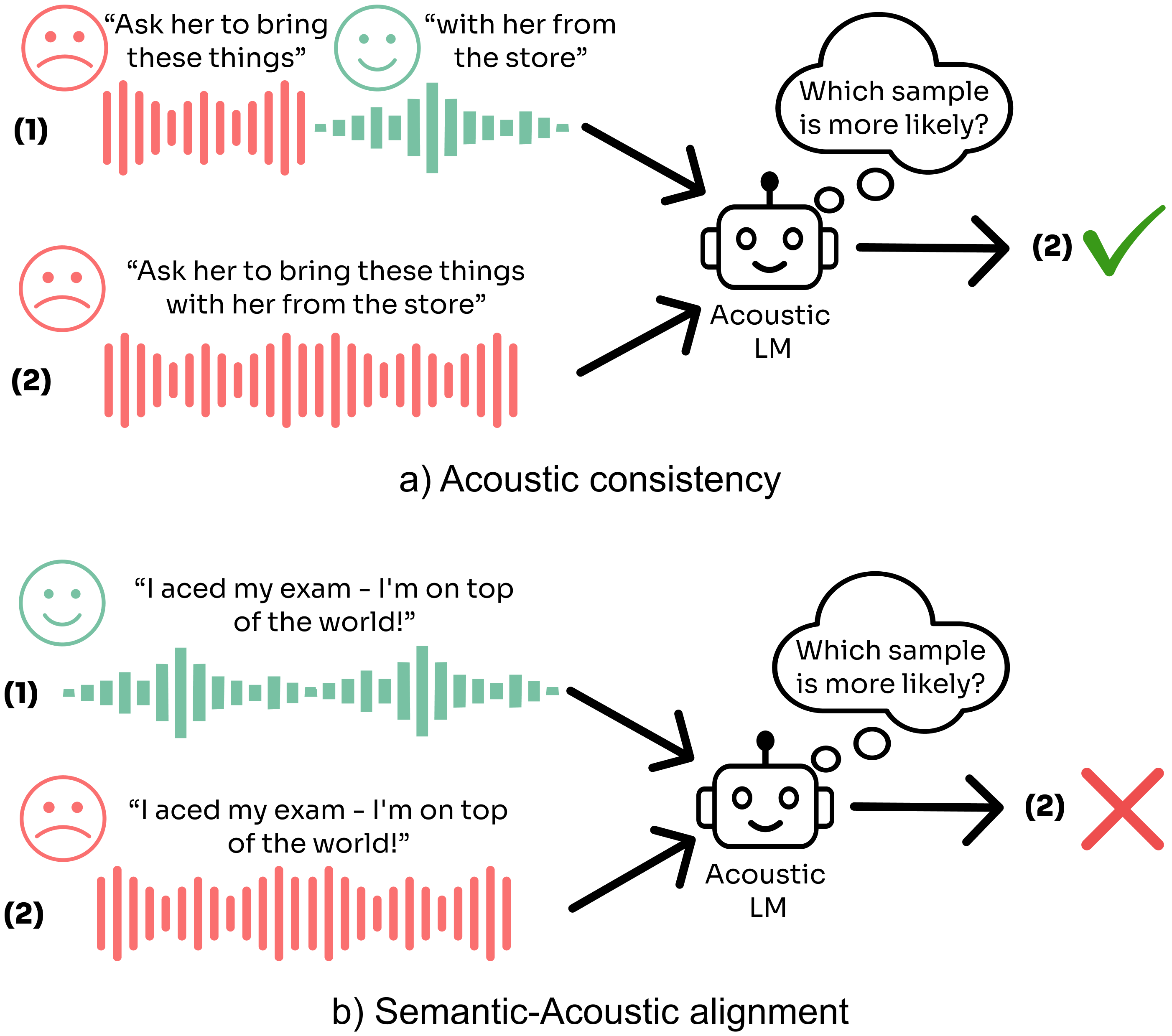}
\caption{A demonstration of \bench~- in which a speech LM is meant to give higher likelihood to real samples. a) shows acoustic consistency, in this case sentiment consistency, where the negative sample changes emotion mid-sentence, and b) shows semantic-acoustic alignment - in this example sentiment alignment.\label{fig:salmon}}
\end{figure}

Parallel work, SD-eval \cite{sd-eval} also proposes a method for evaluating SLMs' ability to address acoustic information. They use text LLMs to generate different text responses suitable for an audio input in a given style (emotion, age, accent and background). They then prompt the SLM with the recording, and compare the generated text response and the ground truth using automated or human metrics. This work studies important speech aspects, but is limited to LMs which output text answers. Furthermore, generation based metrics together with LLM based evaluation are more compute intensive to evaluate thus might become a bottleneck in a model development cycle.

To address the above limitations we introduce \bench, a \emph{Suite for Acoustic Language Model evaluation}. It follows a modelling based evaluation paradigm, i.e. checking that the model assigns higher probability to ``real'' samples, considering a wide variety of acoustic elements. Overall, we present two main tasks: (i) \emph{acoustic consistency} and (ii) \emph{acoustic-semantic alignment}, considering several acoustic aspects. The acoustic consistency benchmark tests whether the model assigns higher likelihood to the original recording, compared with one were an acoustic element (e.g., speaker) changes rapidly in the middle of the sample. The acoustic-semantic alignment metric checks if the model assigns higher scores to samples where the acoustic information matches the spoken content. For instance, the phrase ``I wonder who is at the door?'' is more likely to be heard near a doorbell ring in the background rather than with construction noises. This task is more challenging for SLMs as it requires both semantic text and acoustic understanding, and reasoning capabilities over both information streams jointly. Figure \ref{fig:salmon} provides a visual example.

We evaluate several SLMs using \bench and discuss the insights in Sec. \ref{sec:results}. We show that while humans easily achieve over 90\% on most tasks, SLMs struggle in modelling and identifying basic acoustic inconsistencies. We provide an easy to use evaluation script together with the full benchmarks. We hope \bench will be beneficial for the speech community in guiding future SLMs development towards acoustic modelling together with semantic modelling.
\section{Related Work}\label{sec:rw}
\vspace{-0.3cm}
\newpara{Speech Language Models.}
Much like text language models, speech language models \cite{gslm, audiolm, sashimi} use a next token prediction objective. As speech is continuous by nature, some SLMs operate over continuous speech representations~\cite{algayres2023generative,nachmani2023spoken}, yet the main approach is operating over discrete representations. These can be roughly divided into two main groups: (i) \emph{semantic tokens}~\cite{gslm}, based on applying k-means over latent representations obtained from a pre-trained self-supervised model~\cite{gslm, messica2024nast, mousavi2024should}; (ii) \emph{acoustic tokens}~\cite{encodec,soundstream}, based on neural compression models. As semantic tokens were shown to mostly discard prosody \cite{maimon-adi-2023-speaking, sicherman2023analysing, polyak2021speech}, expressive SLMs augment these speech representations with prosodic features through separate streams of pitch or style tokens either in parallel \cite{pgslm} or flattened into a single stream \cite{spiritlm}.  

In order to use the abilities of text LMs, and to support multi-modal tasks (e.g ASR) some LMs jointly operate over text and speech. Broadly speaking, there are two main approaches: \textit{audio encoding} into the text LM latent \cite{lu2024desta, salmonn}, or \textit{vocabulary expansion}, or hybrids of both \cite{lauragpt, audiopalm}. In \textit{audio encoding} one optimises a mapping from the continuous representation into the continuous LM latent \cite{qwenaudio}, for instance by fine-tuning Whisper \cite{whisper}. This means that we can not directly generate audio, but only prompt a text LM with audio signals. In contrast, \textit{vocabulary expansion} adds speech tokens to the text vocabulary and fine-tunes the LM on both modalities \cite{maiti2024voxtlm}, while some interleave between the modalities~\cite{spiritlm}. This has the benefit of also supporting audio generation and modelling.

\newpara{SLM Evaluation.}
Evaluating SLMs is non-trivial and can broadly be categorised into three types: modelling, generative and task performance. Generative metrics, prompt the SLM with an input and evaluate the output on things such as sentiment consistency \cite{spiritlm} and textual diversity and meaningfulness \cite{gslm}. Task metrics evaluate the ability of the SLM to perform tasks such as ASR or emotion recognition through prompting \cite{audiobench}. Another line of work involves evaluating SLMs as universal speech models via instruction tuning in a zero-shot fashion~\cite{huang2024dynamic, yang2024air}. Finally, modelling metrics check whether the SLM assigns a higher probability to the correct sample compared to others that are unlikely in a specific element \cite{zero_resource, twist}.

Modelling metrics are commonly used to evaluate spoken word related aspects, such as syntax or grammar \cite{zero_resource} or semantic understanding \cite{twist}. They can also be used for prosody related evaluation as in ProsAudit \cite{prosaudit}. Modelling metrics have the benefits of being objective, fast, and easy to compute as they do not require additional models, vocoders, or human studies.

Unlike the above mentioned SLM evaluation benchmarks, \bench measures acoustic elements - speaker identity, sentiment, background noise and room acoustics, using the modelling based approach. It consists of two levels of complexity - one measures the \textit{consistency} of a given acoustic element, whereas the other measures \textit{alignment} between the acoustic details and the semantic, spoken content.

\begin{table*}[t!]
  \caption{Statistics about the different benchmarks within \bench. For the SNR ranges we first randomly sample a range and then sample uniformly within that range to give variation. This, and all other speaker and recording meta-data is recorded per-sample. \label{tab:salmon_data}}
  \centering
  \resizebox{\textwidth}{!}{\begin{tabular}{l|ccccccccc}
    \toprule
      & Datasets & Classes & Sample Length [Sec] & Speakers & SNR ranges [dB] \\

    \midrule
    Sentiment Consistency & Expresso \cite{expresso} & Happy, Sad, Whisper & $5.63 \pm 1.53$ & 4 & - \\
    Speaker Consistency & VCTK \cite{vctk} & 105 speakers & $5.97 \pm 1.70$ & 105 & - \\
    Gender Consistency & VCTK \cite{vctk} & male, female & $5.93 \pm 1.70$ & 104 & - \\
    Background (In-Domain) Consistency & LJ \cite{lj}, FSD50K \cite{fsd50} & 20 Backgrounds & $5.71 \pm 1.42$ & LJ & $(.01, .02), (.1, .2), (1,2), (5,10)$\\
    Background (All) Consistency & LJ \cite{lj}, FSD50K \cite{fsd50} & 20 Backgrounds & $5.61 \pm 1.39$ & LJ & $(.01, .02), (.1, .2), (1,2), (5,10)$\\
    RIR Consistency & LJ, EchoThief\cite{echothief} & 5 RIRs & $5.48 \pm 1.33$ & LJ & - \\
    \midrule
    Sentiment Alignment & AzureTTS~\cite{Azure-TTS}, GPT & Happy, Sad & $4.39 \pm 0.92$ & SaraNeural & - \\
    Background Alignment & FSD50K, AzureTTS, GPT & 20 classes & $7.69 \pm 0.67$ & SaraNeural & - \\
    \bottomrule
  \end{tabular}
  }
\end{table*}

\section{\bench} \label{sec:salmon}
The proposed benchmark is based on the modelling approach described above and evaluates various acoustic aspects, namely speaker identity, sentiment, background and room impulse response. Formally, given an SLM - $M:\text{audio} \to [0,1]$ that assigns a likelihood to a given audio input, for each task $T$ containing samples $s = (s_p, s_n)$ we define the score as:
\begin{equation}
    Score = \frac{1}{|T|}\sum_{(s_p, s_n) \in T}\mathbf{1}_{M(s_p) > M(s_n)},
\end{equation}
i.e. how many times on average the model sets higher likelihood to positive samples over negative ones. On the rare case of identical scores we give 0.5 points, so that an indiscriminate model gets a random score. In practice, we implement the likelihood as follows, 

\begin{equation}
M([w_1, \dots , w_t]) = \frac{1}{t}\sum_{i \le t} p(w_i | w_{i-1}, \dots , w_1),
\end{equation}
but we leave this as a design choice by the LM. Such scoring approach is similar to prior work~\cite{zero_resource}. We split \bench into two types - consistency of the inspected element and its alignment with the semantic content. 

\subsection{Acoustic Consistency}
We wish to evaluate whether SLMs model acoustic elements of the input signal and can detect unnatural acoustic changes. As with other modelling metrics, we evaluate this by comparing the likelihood that an SLM assigns a recording with a given acoustic properties, to the likelihood of the same recording where the some acoustic property changes mid-recording (e.g., background noise, reverberation). See Figure \ref{fig:salmon} (a) for a visual overview.

We achieve this using a dataset with annotated attributes, like speaker, speaking style, and transcriptions. In cases where the attribute to be changed depends on the transcription (e.g., speaker identity), we first apply a forced aligner~\cite{mfa} to get words alignment, then we split the recording by words. In other cases (e.g., background noise change) we simply split the audio by time (and not words). We then take the original recording as the positive sample and mix the first part of the real recording, with the following part from a different recording, with the same text content, as a negative. We give task specific details below and in Table \ref{tab:salmon_data}.

\newpara{Background Consistency.}
For each sample, we randomly choose a speech recording from LJ speech \cite{lj}, and two background noises from FSD50K \cite{fsd50}. We filtered single label examples, manually selected 20 distinct classes, and manually filtered recordings keeping those with clear noises and little silences.
For the positive recording, we merge between the speech and the first background noise. Conversely, for the negative recording, we add the first background to the first part of the speech, and the second background to the rest. This creates a speech recording with varying background noise.

We split this into two sub-tasks: random and in-domain, which vary in background noise sampling for the negative sample. In \emph{random}, we sample two random background noises from the dataset, and in \emph{in-domain} we limit the sampling of the negative background noise to be from the same class as the positive, e.g. the negative recording background changes from one siren to another, making the task more challenging.

\newpara{Sentiment Consistency.}
We use the Expresso dataset~\cite{expresso}. For each sample, we randomly choose 2 recordings corresponding to the same text with different sentiments from happy, sad or whisper. The positive recording is left as is, and the negative is generated by concatenating the first part of the sampled recording, with the second part of another.

\newpara{Speaker Consistency.}
Similar to sentiment consistency, we use VCTK~\cite{vctk} and sample positive recording with the same speaker and negative recording with alternating speakers.

\newpara{Gender Consistency.}
We made an easier version of the above task - called gender consistency, which forces the negative speaker to be from a different gender.

\newpara{Room Impulse Response Consistency.}
Here we measure the ability of the models to grasp the change in acoustic scenes. We use LJ speech \cite{lj} and five diverse RIRs from EchoThief \cite{echothief}. We sample two impulse responses. The positive recording is the speech convolved with the first impulse response. We construct the negative by convolving the first half of the speech with the first RIR, the second half with the second RIR and concatenating the two. This procedure results in two recordings of the same speech, one in a given room, and in the other the room is switched mid-sample.

\begin{table*}[t!]
      \caption{Comparison of leading SLMs on the \bench benchmark. Bg (domain) and Bg (rand.) stands for background noise sampled from the same domain (i.e. class) or at random, respectively. \label{tab:salmon_eval}}
  \centering
  \resizebox{\textwidth}{!}{\begin{tabular}{l|cccccc|cc|c}
    \toprule
    
      & \multicolumn{6}{|c|}{\textbf{Acoustic Consistency}} & \multicolumn{2}{|c|}{\textbf{Semantic-Acoustic Alignment}} & \textbf{Spoken Content}\\

      & Sentiment $\uparrow$ & Speaker $\uparrow$ & Gender $\uparrow$ & Bg (domain) $\uparrow$ & Bg (rand.) $\uparrow$ & Room $\uparrow$ & Sentiment $\uparrow$ & Background $\uparrow$ & sWUGGY $\uparrow$ \cite{zero_resource} \\

    \midrule
    TWIST 350M\cite{twist} & 59.0 & 69.5 & 68.0 & 54.0 & 61.5 & 59.0 & 51.5 & 56.5 & 80.7\\
    TWIST 1.3B\cite{twist} & 61.5 & 69.0 & 69.5 & 55.5 & 60.5 & 59.0 & 53.0 & 56.5 & 81.0\\
    TWIST 7B\cite{twist}   & 61.5    & 71.0    & 70.0   & 55.0    & 60.5    & 62.0    & 51.5    & 54.5    & \textbf{82.8}\\

    LAST 350M\cite{last}   & 64.0 & 63.0 & 70.5 & 55.5 & 60.5 & 61.0 & 51.5 & 54.5 & 73.5\\
    LAST 1.3B\cite{last}   & 65.0 & 64.5 & 68.5 & 56.0 & 61.0 & \textbf{62.5} & 53.5 & 53.0 & 73.6\\

    pGSLM\cite{pgslm}      & 40.5 & \textbf{83.0} & \textbf{88.5} & 
    \textbf{57.0} & \textbf{66.0} & 53.5 & 55.5 & 53.5 & 74.1\\

    Spirit LM\cite{spiritlm} & 54.5 & 69.5 & 67.0 & 
    53.5 & 55.5 & 54.5 & 48.0 & 51.5 & 75.5\\
    Spirit LM Expr.\cite{spiritlm}& \textbf{73.5} & 81.0 & 85.0 & 
    56.0 & 64.0 & 54.5 & 52.0 & \textbf{59.5} & 72.7\\

    ASR+LLM\cite{whisper, llama2} & 53.5 & 52.2& 52.0 & 50.0 & 50.5 & 53.2 & 50.5 & 51.5 & 77.0\\
    \midrule
    Human                        & \textbf{\underline{97.2}} & \textbf{\underline{91.5}} & \textbf{\underline{98.6}} & \textbf{\underline{83.1}} & \textbf{\underline{88.7}} & \textbf{\underline{94.4}} & \textbf{\underline{93.3}} & \textbf{\underline{95.8}} & - \\
    \bottomrule
  \end{tabular}
  }
\end{table*}

\subsection{Acoustic-Semantic Alignment}
This measures if an acoustic element such as background noise is aligned with the text. For instance, given a recording of the phrase ``I love the smooth sound of a piano" we expect a recording with piano music to be more likely compared to construction noises. Figure \ref{fig:salmon} (b) shows a visual example.

We construct this benchmark by using GPT4~\cite{GPT-4o} to create texts that someone would likely say under a condition, i.e. hearing a sound or in a given sentiment. We manually filter these texts samples to leave only clear samples. We then use Azure text-to-speech \cite{Azure-TTS} to synthesise the text.

\newpara{Background Alignment.}
This task focuses on evaluating the alignment between background noise and the speech content. For each background noise class, from the filtered FSD50K we use in \textit{background consistency}, we generated 20 texts using GPT4~\cite{GPT-4o} corresponding to speech one is likely to hear with this class of background noise. As mentioned, we manually filter these texts to keep clear examples. We synthesize the speech using Azure TTS \cite{Azure-TTS}. For the positive recording, we sample a random background noise from the positive class and pre-pend it to the synthesized speech. We generate the negative recording by sampling a random background noise from a different background class and merging it with the speech corresponding to the positive background class.

\newpara{Sentiment Alignment.}
Lastly, we evaluate SLMs capability to model the relation between spoken content, and speech sentiment. We use GPT4~\cite{GPT-4o} to generate ${\sim}200$ sentences which someone would say in a cheerful or sad sentiment. For each sample, we randomly picked a sentence and generated emotional speech for both sentiments with Azure expressive TTS. The positive sample is that where the text sentiment matches the TTS sentiment. We manually filter the resulting samples rather aggressively as the quality of the TTS is limited, especially in conflicting sentiment situations. We aim to remove any samples where the cheerful might come off as emphatic or stressed, and the sad may come off as emotional or touched, causing ambiguity. We also try to remove samples with noticeable synthesis artifacts. This manual process is inherently limited and subjective so we evaluate the effectiveness by human evaluation in section \ref{sec:baselines}.
\section{Baselines}\label{sec:baselines}
We evaluate the performance of popular SLMs on the different parts of \bench. Through this we evaluate the impact of different model aspects, such as number of parameters and expressive modelling approaches.

We use TWIST~\cite{twist}, which uses a pre-trained text-LM as an initialisation for SLM training over HuBERT~\cite{hubert} units. They show this noticeably improves semantic metrics such as sWUGGY, and we wish to evaluate how they perform on acoustic metrics. We also utilise the existence of the three model sizes (350M, 1.3B, 7B) to see whether this makes a difference, or if the use of HuBERT units only, limits the performance. 

We additionally explore LAST \cite{last} as a different speech tokeniser. It is a recently proposed tokeniser guided by a pre-trained text LM. We note, that in this setup, LAST can be considered a TWIST model with a different tokeniser.

We further wish to evaluate an explicitly expressive SLM baseline, and choose pGSLM \cite{pgslm}. pGSLM trains an SLM over three streams: de-duplicated HuBERT units, unit durations, and a quantised pitch contour. This is meant to allow prosody modelling, which is a big part of acoustics evaluated in \bench.

To evaluate the impact of integrating text into SLMs, on their acoustic modelling abilities we also evaluate Spirit-LM\cite{spiritlm}. Spirit-LM continues pre-training of a text-LM, on interleaved text-speech data by expanding the vocabulary of the LM with speech units. It comes in both base and expressive versions, with the expressive version augmenting HuBERT with pitch and style tokens meant to capture prosody. We use the official pre-trained model released by the authors\footnote{\url{https://github.com/facebookresearch/spiritlm}}, the authors claim that this is a slight variation from the original but should give similar results.

\section{Human Evaluation} 
Next, to both validate \bench and provide an upper bound on model performance, we conduct a human study. In it, we measure humans' ability to rate the positive samples as more likely than the negative samples. While this task is trivial to people when ``prompted'' with the exact task question - e.g. ``In which sample does the text sentiment best match the speech sentiment?'', we wanted to assert that humans agree with the labels even with the general question ``Which sample is more likely?''. This bundles in other elements which could impact likelihood such as TTS artifacts, speakers with strong accents etc, thus perfectly emulating the task the SLM faces. We select 20 samples from each benchmark, and each is annotated by at least three annotators fluent in English.

\section{Results}\label{sec:results}
\vspace{-0.25cm}
\newpara{Consistency metrics.}
In analysing the more simple acoustic consistency tasks in Table \ref{tab:salmon_eval}, we see that Gender consistency is the most trivial. The best model on this task, pGSLM, achieves high performance of 88.5 which is still shy of human performance (98.6). Likewise, on the slightly more challenging, generalised version of this - speaker consistency, with pGSLM reaching 83.0. 

We observe that non-expressive methods which operate over HuBERT units, such as TWIST, perform notably worse even when almost two orders of magnitude larger in parameter count. However, it is interesting to note that at the LM level, they can discern the speaker to an extent (as shown by the better than random performance), even if their vocoder might be single speaker.

For other consistency tasks, specifically - Sentiment, Impulse Response (Room), and Background - the performance is better than random for some models. However, even for the best performing models the results are far from human performance. We note that increasing the model size in TWIST and LAST has little to no effect on model performance. 

We observe that non-expressive Spirit-LM, performs comparably or marginally worse than TWIST 7B, of the same size, across all tasks. This could indicate that including text as part of the LM, in interleaving fashion, improves semantic abilities, but does not benefit the acoustic abilities. Furthermore, the expressive version which does model prosodic information performs comparably to pGSLM across most tasks, strengthening this claim. The improved performance in the sentiment consistency task could derive from better style tokens, more data or a larger model size.

We believe that many of these results could derive from lack of expressive modelling (i.e. HuBERT tokens only) or from lack of training diversity. For instance, HuBERT units are likely ill-equipped to model background noise, but this could also be true for pitch trackers in pGSLM which are aimed at speech. Conversely, we expect pitch to differ in emotional speech, making the Sentiment Consistency task reasonable for pGSLM. Furthermore, the training data of these models is largely based on audio-books and podcasts which are highly unlikely to contain reverberation or background noise, as studio recording settings aim to remove these as much as possible.

\newpara{Alignment metrics.}
The results are summarized in Table~\ref{tab:salmon_eval}. When considering the results on the semantic-acoustic alignment tasks, we observe that no baseline achieves substantial improvements over the random baseline (59.5 maximum score, and 56.5 second best). This is despite the fact that this task is trivial to humans which achieve 93.3 and 95.8 on \textit{sentiment alignment} and \textit{background alignment} respectively.

Interestingly, while expressive Spirit-LM, does manage to model sentiment to some extent (as we can see from the sentiment consistency task), it achieves near random performance on the responding alignment task. When considering sWUGGY scores, and the results reported in the original paper, the model contains textual modelling capabilities. Nevertheless, the ability to jointly model both text and acoustic content and reason over them is still lacking. These tasks remain an interesting challenging task for acoustic aware SLMs.

We also note that a cascaded pipeline of ASR followed by a text-LM achieves practically random performance across all tasks. While this is intuitive because the text is identical in both samples, we show that the ASR does not leak additional information. Furthermore, for many semantic metrics this baseline is a leading option thus highlighting the importance of acoustic metric evaluation to steer research efforts at methods which jointly improve acoustic modelling and semantic modelling. We also observe that models which have higher sWUGGY scores, such as TWIST, do not necessarily outperform methods like pGSLM in acoustic evaluation, thus it is important to focus and evaluate both metrics.
\vspace{0.1cm}
\section{Conclusion}
We introduce \bench, a suite for evaluating acoustic LMs on many acoustic aspects, namely: gender, speaker, background noise, sentiment, and room impulse response. We achieve this through a modelling based metric which is objective and fast to compute. We evaluate several popular SLMs, and human raters, on the \bench benchmark and show that current models are far behind human performance on the evaluated tasks. We hope that publishing this benchmark and sample generation pipeline will progress the development of acoustic aware SLMs. 

\newpara{Acknowledgements.} This research work was supported by ISF grant 2049/22.

\bibliographystyle{IEEEbib}
\bibliography{refs}

\end{document}